\documentclass[
 reprint,
nofootinbib,
 amsmath,amssymb,
 aps,
 superscriptaddress,
]{revtex4-2}

\usepackage{graphicx}
\usepackage{dcolumn}
\usepackage{bm}
\usepackage[caption=false]{subfig}
\usepackage{multirow}
\usepackage{rotating}
\usepackage{array}
\usepackage{url}
\usepackage{braket}
\usepackage{hyperref}
\usepackage{tikz}
\usetikzlibrary{quantikz2}
\usepackage{accents}

\begin{document}

\preprint{APS/123-QED}

\title{Variational Quantum Machine Learning with Quantum Error Detection}
\author{Eromanga Adermann}
\email{eromanga.adermann@csiro.au}
\affiliation{Data61, CSIRO, 26 Pembroke Rd, Marsfield, NSW, 2122, Australia}

\author{Hajime Suzuki}
\affiliation{Data61, CSIRO, 26 Pembroke Rd, Marsfield, NSW, 2122, Australia}

\author{Muhammad Usman}
\affiliation{Data61, CSIRO, Research Way, Clayton, Vic, 3168, Australia}
\affiliation{School of Physics, The University of Melbourne, Parkville, Vic, 3010, Australia}

\date{\today}

\begin{abstract}
Quantum machine learning (QML) is an emerging field that promises advantages such as faster training, improved reliability and superior feature extraction over classical counterparts. However, its implementation on quantum hardware is challenging due to the noise inherent in these systems, necessitating the use of quantum error correction (QEC) codes. Current QML research remains primarily theoretical, often assuming noise-free environments and offering little insight into the integration of QEC with QML implementations. To address this, we investigate the performance of a simple, parity-classifying Variational Quantum Classifier (VQC) implemented with the [[4,2,2]] error-detecting stabiliser code in a simulated noisy environment, marking the first study into the implementation of a QML algorithm with a QEC code. We invoke ancilla qubits to logically encode rotation gates, and classically simulate the logically-encoded VQC under two simple noise models representing gate noise and environmental noise. We demonstrate that the stabiliser code improves the training accuracy at convergence compared to noisy implementations without QEC. However, we find that the effectiveness and reliability of error detection is contingent upon keeping the ancilla qubit error rates below a specific threshold, due to the propagation of ancilla errors to the physical qubits. Our results provide an important insight: for QML implementations with QEC codes that both require ancilla qubits for logical rotations and cannot fully correct errors propagated between ancilla and physical qubits, the maximum achievable accuracy of the QML model is limited. This highlights the need for additional error correction or mitigation strategies to support the practical implementation of QML algorithms with QEC on quantum devices. 
\end{abstract}

\maketitle

\section{Introduction}
The integration of quantum computing with machine learning is predicted to address the increasing demand for computational power and efficiency across a wide range of complex tasks within the field of machine learning~\cite{Biamonte_2017, Cong_2019, Havl_ek_2019, Schuld_2019, Beer_2020}. QML algorithms aim to exploit the fundamental quantum phenomena of superposition and entanglement that are unavailable to classical machine learning algorithms. By harnessing these unique properties of quantum mechanics, QML offers the potential for computational speed-ups, improved model performance compared to their classical counterparts for certain classes of problems~\cite{Lloyd2013QuantumAF, PhysRevLett.113.130503, 2014NatPh..10..631L}, including enhanced pattern recognition through the use of quantum feature maps that lead to quantum speed-ups~\cite{Liu_2021, Wang_2025}, and greater robustness to adversarial attacks exhibited by QML models~\cite{wu2023radiosignalclassificationadversarially, West_2023, West_2024, dowling2024adversarialrobustnessguaranteesquantum, Khatun_2024}. 

Given that the vast majority of QML experiments have thus far been conducted theoretically in ideal classical simulation environments, it is unclear whether the advantages predicted so far will be retained when implementing QML algorithms on realistic quantum hardware, such as on Noisy Intermediate Scale Quantum (NISQ) devices~\cite{Preskill_2018, Bharti_2022}. As for all quantum algorithms, one of the biggest obstacles to the practical implementation of QML on quantum devices is vulnerability to noise, which can cause errors and distort computations, leading to meaningless outputs. 

Numerous strategies have been developed to suppress and mitigate noise with the goal of achieving quantum advantage prior to the advent of fully fault-tolerant quantum systems, including approximate algorithms such as the Quantum Approximate Optimisation Algorithm~\cite{farhi2014quantumapproximateoptimizationalgorithm}, heuristic approaches~\cite{Biamonte_2017} and approximate amplitude encoding~\cite{West_2024}, all of which reduce circuit depths and minimise noise accumulation. Noise-induced errors can also be reduced using virtual distillation~\cite{PhysRevX.11.041036, Karim_2024}, which suppresses errors by combining multiple noisy copies of a quantum state, and dynamical decoupling~\cite{Lidar_2014, PhysRevApplied.20.064027, Ji_2024}, which preserves coherence during computation by applying carefully-timed pulses to qubits. While these methods may be applied to QML, and while some have already proven to be effective in implementations of QML on a quantum device \cite{West_2024}, achieving scalability and fault tolerance for quantum algorithms ultimately requires the adoption of Quantum Error Correction (QEC) codes~\cite{Gottesman_1998}. 

Early QEC research focused on stabiliser codes \cite{Calderbank_1996, Steane_1996, shor1997faulttolerantquantumcomputation}, and established the Threshold Theorem, which states that fault-tolerant quantum computation is possible provided physical error rates in quantum hardware remain below a finite threshold~\cite{aharonov1996faulttolerantquantumcomputation, Kitaev1997}. Since then, QEC has expanded to include diverse codes tailored for different advantages and environments, such as surface codes~\cite{Dennis_2002}, Bacon-Shor codes~\cite{Bacon_2006} and 3D color codes~\cite{Bombin_2006}. Very recently we have begun to see experimental verification of the effectiveness of these codes, such as in the landmark demonstration of the surface code operating below its critical threshold on superconducting processors~\cite{acharya2024quantumerrorcorrectionsurface}. However, the present challenge with QEC codes is their high resource overhead, with practical implementations of useful algorithms potentially requiring hundreds of thousands of qubits~\cite{Kivlichan_2020}. 

As a result, the theoretical and experimental implementation of QEC codes is still in its early stages and is largely limited to the smallest codes or the simplest operations. QEC codes have been successfully applied to a small number of quantum operations in both theoretical and experimental quantum environments. In particular, the [[4,2,2]] stabiliser code has seen widespread use in recent years across a range of applications, including magic state preparation~\cite{Gupta_2024}, quantum chemistry~\cite{vandam2024endtoendquantumsimulationchemical, bedalov2024faulttolerantoperationmaterialsscience}, Variational Quantum Eigensolvers~\cite{Urbanek_2020, zhang2022comparativeanalysiserrormitigation, gowrishankar2025logicalerrorrates422encoded}, and the implementation of diverse quantum circuits~\cite{cane2021experimentalcharacterizationfaulttolerantcircuits, Sun2022, reichardt2024logicalcomputationdemonstratedneutral}. Codes with enhanced error correction capabilities and greater resource requirements, including the Steane code~\cite{PhysRevLett.77.793, Steane1996-bw}, repetition code~\cite{Kelly_2015}, Shor code~\cite{PhysRevA.52.R2493}, Bacon-Shor code and surface code, have primarily been applied to error correction on a single logical qubit~\cite{hilder2021faulttolerantparityreadoutshuttlingbased, egan2021faulttolerantoperationquantumerrorcorrection, Zhao_2022, Acharya2023} or to the implementation of simple operations on 1-2 logical qubits~\cite{, Bluvstein_2023, Het_nyi_2024, paetznick2024demonstrationlogicalqubitsrepeated, Kim_2025}, though notably the Steane Code was recently used to logically encode three-qubit circuits for the quantum Fourier Transform~\cite{mayer2024benchmarkinglogicalthreequbitquantum}. Despite these recent advancements, no experiment or theoretical simulation has yet demonstrated the application of a QEC code to a QML problem. 
 
We present the first study on the implementation of a QML algorithm with a QEC code. Specifically, we implement a simple Variational Quantum Classifier (VQC) with the [[4,2,2]] stabiliser code. We selected the simplest stabiliser code and QML algorithm in order to minimise resource overhead in our experiments, which nevertheless requires 20 qubits to implement five rounds of syndrome extractions. In Sections ~\ref{Encoding} and~\ref{sims}, we explain how we logically encoded the VQC according to the [[4,2,2]] code, introduce the noise models we used to simulate realistic noisy environments, and detail the parameters used in our simulations. We then present and discuss the results of our analyses in Section~\ref{Results}, showing that a threshold error rate for ancilla qubits exists, such that above this threshold the QEC code is no longer capable of ensuring good training accuracies and state fidelities. We conclude in Section~\ref{conclusion} with a discussion of the implications of our findings for the practical implementation of QML algorithms on NISQ and fully fault-tolerant systems. 

\section{Experimental set-up}\label{Encoding}
\begin{figure*}
\centering
\begin{quantikz}
\ket{0} & \phantomgate{0} & \gate{X}\gategroup[2,steps=1,style={dashed, rounded corners, inner sep=6pt}]{Basis Encoding} & \phantomgate{0} & \gate{R_{X}(\theta)}\gategroup[2,steps=4,style={dashed,rounded corners,inner sep=6pt}]{Variational Component}  & \gate{R_{Z}(\theta)} & \ctrl{1} & \gate{R_{Y}(\theta)} & \gate{\sigma_z} & \meter{} \\
\ket{0} & \phantomgate{0} & \phantomgate{0} & \phantomgate{0} & \gate{R_{X}(\theta)}& \gate{R_{Z}(\theta)} & \targ{} & \gate{R_{Y}(\theta)} &\phantomgate{0} & \phantomgate{0}
\end{quantikz}
\caption{The Variational Quantum Classifier (VQC) with an example input state of $\ket{10}$ and rotational parameter $\theta$.}
\label{VQC}
\end{figure*}
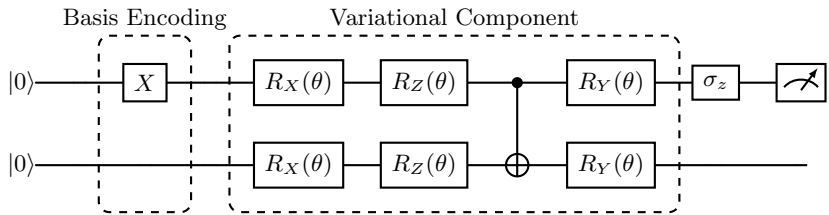

We chose a very simple 2-qubit VQC for our experiments (displayed in Figure~\ref{VQC}), in order to minimise both resource overhead and computational time required to run the simulations. The VQC takes two qubits encoded in the basis encoding as input, and classifies their parity through measurement of the first qubit in the $Z$ basis. The quantity measured is the expectation value over $1000$ shots. We use only one rotational parameter, $\theta$, to train the classifier, as any more than one leads to overfitting. The classifier is able to reach an accuracy of $1.0$ within 100 training iterations. 

\subsection{Logical Encoding}
We chose the [[4,2,2]] 4-qubit Calderbank-Shor-Steane (CSS) stabiliser code for our encoding and error detection, as it is the simplest stabiliser code that protects against $X$ and $Z$ single-qubit errors \cite{Vaidman_1996}. It encodes 2 logical qubits using 4 physical qubits, and can only facilitate detection (not correction) of single-qubit errors. As with all stabiliser codes, errors are detected by taking measurements of ancilla qubits after applying stabilisers to the physical qubits, known as syndrome extraction. 

We used the following mapping to encode 2 logical qubits with 4 physical qubits: 
\begin{align}
    \ket{00}_{L} &= \frac{1}{\sqrt{2}}(\ket{0000} + \ket{1111})\\
    \ket{01}_{L} &= \frac{1}{\sqrt{2}}(\ket{0011} + \ket{1100})\\
    \ket{10}_{L} &= \frac{1}{\sqrt{2}}(\ket{0101} + \ket{1010})\\
    \ket{11}_{L} &= \frac{1}{\sqrt{2}}(\ket{0110} + \ket{1001}), 
\end{align}
where the left-hand side represents the four different 2-logical qubit states (indicated by the subscript $L$), and the right-hand side represents the physical qubit states. 

With these definitions, the CNOT gate in the VQC is logically encoded by a SWAP gate between the first two physical qubits. The logical encoding for the rotation gates require additional ancilla qubits (one ancilla per rotation gate to be encoded), where the ancilla qubits undergo the rotation instead of the physical qubits. This process generally requires a series of CNOT gates before and after the application of each rotation gate, which entangle the physical (and logical) qubits with the ancilla qubits. 

In our logical encoding, the quantum state of the full qubit system, including ancilla qubits, is always of the form: 
\begin{equation}
\label{superposition_state}
    \ket{\Psi} = \sum_{i=0}^{3} c_i \ket{\psi_i}_L \bigotimes_{j=0}^{n-1}\,\ket{\phi_{i}}_{a_j},
\end{equation}
where $\ket{\psi_i}_L$ represents each of the four possible logical basis states, $c_i$ is the complex coefficient associated with the $i$-th logical basis state, $\ket{\phi_i}_{a_j}$ represents the state of the $j$-th ancilla qubit, $a_j$, associated with the $i$-th logical basis state, and $n$ is the total number of ancilla qubits that have been introduced into the system. 

To illustrate, the full quantum state after the first two $R_{X}$ rotations is given by: 
\begin{align}
    \ket{\Psi} &= -i\mathrm{cs}\ket{00}_L \ket{0}_{a_1}\ket{0}_{a_2} + \mathrm{c^{2}}\ket{01}_L\ket{0}_{a_1}\ket{1}_{a_2} \nonumber\\
    &- \mathrm{s^{2}}\ket{10}_L\ket{1}_{a_1}\ket{0}_{a_2} - i\mathrm{sc}\ket{11}_L\ket{1}_{a_1}\ket{1}_{a_2},
\end{align}
where $c$ and $s$ are short-hand notations for cos($\theta$) and sin($\theta$), the logical states are subscripted with $L$ and the ancilla states are subscripted with $a_j$. There are two ancilla states ($a_1$, $a_2$) invoked because two rotations have occurred. By the end of a logical operation, the ancilla qubits in each term reflect the logical state within the same term. 

The steps we used to logically perform the double $R_Y$ and $R_X$ gates are outlined below:
\begin{enumerate}
\item \textbf{New ancilla initiation:} If no rotations have been performed in previous steps, initiate two new ancilla qubits to match the initial input logical state to the circuit. For example, if the initial input logical state is $\ket{01}_L$, then the two new ancilla qubits, $a_j$ and $a_{j+1}$, should respectively be in the states $\ket{0}$ and $\ket{1}$. If double rotations have been performed in previous steps, invoke the new ancilla qubits in the $\ket{0}$ state, then match the new ancilla states to the previous two ancilla states that were invoked to perform the last double rotation, by applying CNOT gates to the new ancilla qubits controlled by the two previous ancilla qubits. 
\item \textbf{Change previous ancilla states:} If double rotations have been performed in previous steps, apply CNOT gates to each of the previously invoked ancilla qubits, controlled by the new ancilla qubits. For example, for newly invoked ancilla qubits $a_j$ and $a_{j+1}$, we apply CNOT gates to any previous ancillas $a_{j-2}$, $a_{j-4}$, ..., $a_0$ and $a_{j-1}$, $a_{j-3}$, ..., $a_1$ controlled by $a_j$ and $a_{j+1}$ respectively. 
\item \textbf{Change logical state:} Apply CNOT gates to physical qubits $q_1$ and $q_3$ controlled by newly initiated ancilla $a_j$, and CNOT gates to $q_2$ and $q_3$ controlled by newly initiated ancilla $a_{j+1}$. 
\item \textbf{Apply rotation gate:} Apply the relevant rotation gates to each of the 2 ancilla qubits. 
\item \textbf{Undo logical state change:} Apply the same set of CNOT operations targeting the physical qubits and controlled by the 2 ancilla qubits as was performed in Step 3. 
\item \textbf{Undo previous ancilla state change:} Apply the same set of CNOT operations as applied in Step 2, targeting the ancilla qubits invoked for previous rotations, and controlled by the newest ancilla states. 
\end{enumerate}

\begin{figure*}[ht]
    \centering
    \subfloat[Logical $R_{X}$ rotations]{
        \begin{quantikz}
            \lstick{$\ket{q_0}$} & \qw & \qw & \qw & \qw & \qw & \qw & \qw & \qw & \qw & \qw\\
            \lstick{$\ket{q_1}$} & \targ{} & \qw & \qw & \qw & \qw & \targ{} & \qw & \qw & \qw & \qw\\
            \lstick{$\ket{q_2}$} & \qw & \qw  & \targ{} & \qw & \qw & \qw & \qw & \targ{} & \qw & \qw\\
            \lstick{$\ket{q_3}$} & \qw & \targ{} & \qw & \targ{} & \qw & \qw & \targ{} & \qw & \targ{} & \qw\\
            \lstick{$\ket{a_0}$} & \ctrl{-3} & \ctrl{-1}  & \qw & \qw & \gate{R_{X}(\theta)} & \ctrl{-3} & \ctrl{-1} & \qw & \qw & \qw\\
            \lstick{$\ket{a_1}$} & \qw & \qw & \ctrl{-3} & \ctrl{-2} & \gate{R_{X}(\theta)} & \qw & \qw & \ctrl{-3} & \ctrl{-2} & \qw
        \end{quantikz}
        \label{fig:sub1}
    }
    \hspace{2em} 
    \subfloat[Logical $R_{Z}$ rotations]{
        \begin{quantikz}
            \lstick{$\ket{a_0}$} & \ctrl{2} & \qw & \qw & \qw \\
            \lstick{$\ket{a_1}$} & \qw  & \ctrl{2} & \qw & \qw\\
            \lstick{$\ket{a_2}$} & \targ{} & \qw  & \gate{R_{Z}(\theta)} & \qw\\
            \lstick{$\ket{a_3}$} & \qw  & \targ{} & \gate{R_{Z}(\theta)} & \qw\\
        \end{quantikz}
        \label{fig:sub2}
    }
    \hspace{2em} 
    \subfloat[Logical $R_{Y}$ rotations]{
        \begin{quantikz}
            \lstick{$\ket{q_0}$} & \qw & \qw & \qw & \qw & \qw & \qw & \qw & \qw & \qw & \qw & \qw & \qw & \qw & \qw & \qw & \qw & \qw & \qw & \qw & \qw\\
            \lstick{$\ket{q_1}$} & \qw & \qw & \targ{} & \qw & \qw & \qw & \qw & \qw & \qw & \qw & \qw & \targ{} & \qw & \qw & \qw & \qw & \qw & \qw & \qw & \qw\\
            \lstick{$\ket{q_2}$} & \qw & \qw & \qw & \qw  & \targ{} & \qw & \qw & \qw & \qw & \qw & \qw & \qw & \qw  & \targ{} & \qw & \qw & \qw & \qw & \qw & \qw\\
            \lstick{$\ket{q_3}$} & \qw & \qw & \qw & \targ{} & \qw & \targ{} & \qw & \qw & \qw & \qw & \qw & \qw & \targ{} & \qw & \targ{} & \qw & \qw & \qw & \qw & \qw\\
            \lstick{$\ket{a_0}$} & \qw & \qw & \qw & \qw  & \qw & \qw & \targ{} & \qw & \qw & \qw & \qw & \qw & \qw  & \qw & \qw & \targ{} & \qw & \qw & \qw & \qw\\
            \lstick{$\ket{a_1}$} & \qw & \qw & \qw & \qw & \qw & \qw & \qw & \qw & \targ{} & \qw & \qw & \qw & \qw & \qw & \qw & \qw & \qw & \targ{} & \qw & \qw\\
            \lstick{$\ket{a_2}$} & \ctrl{2} & \qw & \qw & \qw  & \qw & \qw & \qw & \targ{} & \qw & \qw & \qw & \qw & \qw  & \qw & \qw & \qw & \targ{} & \qw & \qw & \qw\\
            \lstick{$\ket{a_3}$} & \qw & \ctrl{2} & \qw & \qw & \qw & \qw & \qw & \qw & \qw & \targ{} & \qw & \qw & \qw & \qw & \qw & \qw & \qw & \qw & \targ{} & \qw\\
            \lstick{$\ket{a_4}$} & \targ{} & \qw & \ctrl{-7} & \ctrl{-5}  & \qw & \qw & \ctrl{-4} & \ctrl{-2} & \qw & \qw & \gate{R_{Y}(\theta)} & \ctrl{-7} & \ctrl{-5} & \qw & \qw & \ctrl{-4} & \ctrl{-2} & \qw & \qw & \qw\\
            \lstick{$\ket{a_5}$} & \qw & \targ{} & \qw & \qw & \ctrl{-7} & \ctrl{-6} & \qw & \qw & \ctrl{-4} & \ctrl{-2} & \gate{R_{Y}(\theta)} & \qw & \qw & \ctrl{-7} & \ctrl{-6} & \qw & \qw & \ctrl{-4} & \ctrl{-2} & \qw
        \end{quantikz}
        \label{fig:sub3}
    }
    \caption{Logical rotation gates implemented with the [[4,2,2]] encoding: (a) logical $R_{X}(\theta)$, (b) logical $R_{Z}(\theta)$ and (c) logical $R_{Y}(\theta)$ rotations. In each subfigure, the physical qubits are denoted by $\ket{q_i}$ and the ancilla qubits are denoted by $\ket{a_i}$.}
    \label{logical_encoding}
\end{figure*}
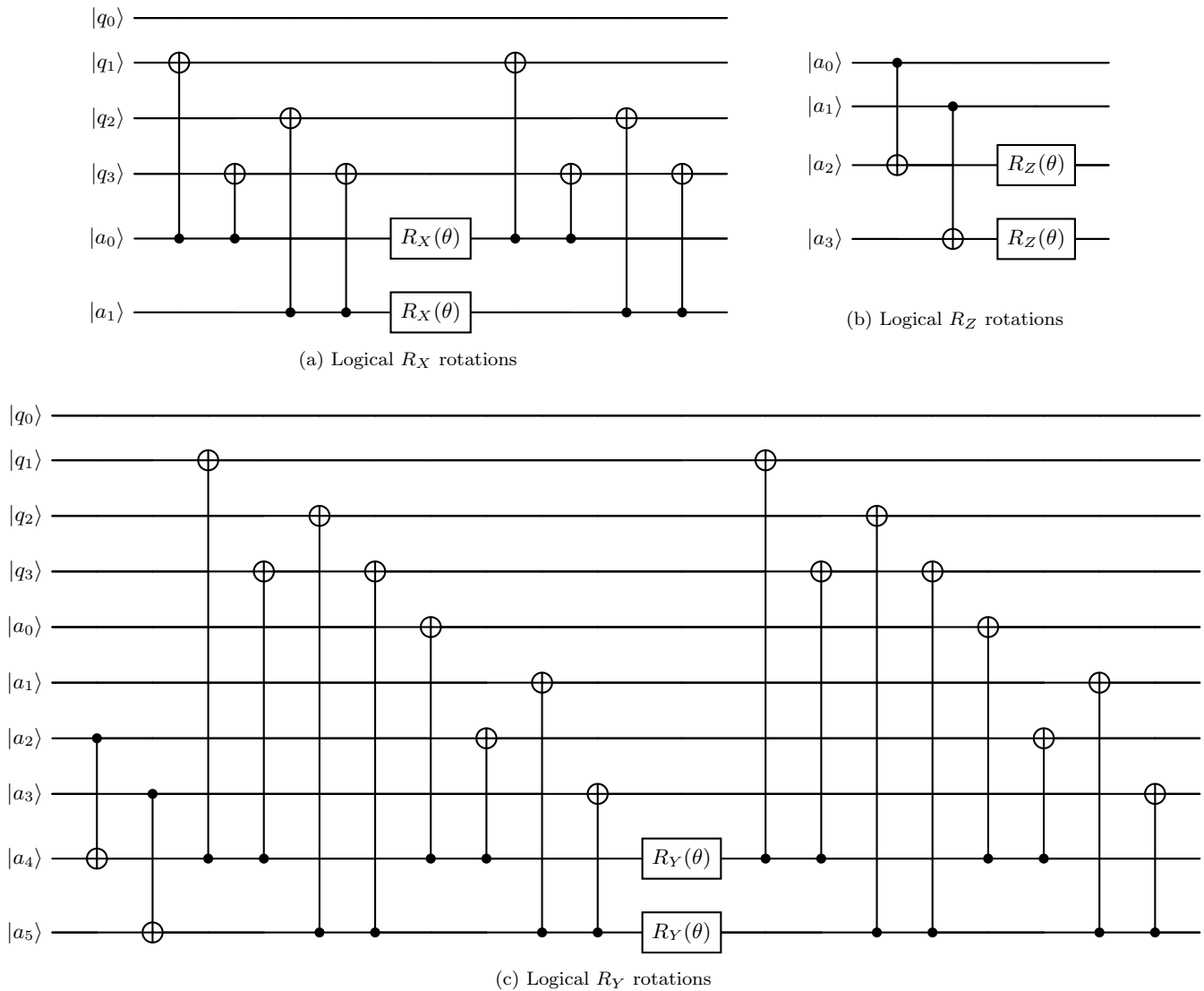

The steps for implementing the $R_Z$ gates are much simpler and do not require as many CNOT gates between physical and ancilla qubits. We only match the newly introduced ancilla qubits to the previous two ancilla qubits, then apply the $R_Z$ rotation gate to each new ancilla qubit. In Figure~\ref{logical_encoding}, we show the full set of operations we used to perform the logical equivalent of the double $R_X$, $R_Z$ and $R_Y$ gates shown in Figure~\ref{VQC}. For the logical CNOT gate, we applied additional CNOT gates after the SWAP gate to ensure matching between the ancilla and logical states in each term of the full quantum state of the system. 

The above steps can also be used to logically implement single qubit rotations, in which case we only need to introduce one ancilla qubit each time. Logically rotating a logical qubit generally requires initiating a new ancilla qubit, transforming the states of the ancillas that were used to logically rotate the qubit in previous steps, transforming the qubit state itself, followed by applying the relevant rotation gate to the newly initiated ancilla, and finally re-applying the same operations to the ancillas and logical qubit that were applied before the rotation. This approach to logically performing rotation gates can be generalised to circuits with any number of rotation gates. 

The logical circuit does not allow the direct $Z$ basis measurement of the first logical qubit, so we conducted the logical equivalent by measuring the probability distribution across the 16 states spanned by the four physical qubits, in the $Z$ basis. Using these probabilities, we calculated the equivalent probability distribution for the four logical 2-qubit states, from which we determined the expectation value of the $Z$ basis measurement of the first logical qubit. 

\subsection{Noise and Error models}
We considered two types of incoherent noise that can occur in a quantum circuit: probabilistic gate noise and environmental noise. Gate noise can arise from imperfections in hardware or control signals, qubit cross-talk during multi-qubit operations, and otherwise non-ideal behaviour of the qubit system whenever a gate is implemented. Environmental noise typically consists of noise that is external to the qubit system seeping in, including stray electromagnetic (EM) fields, photons, and mechanical vibrations. These types of noise are usually modelled by their impact on qubits; namely, thermal relaxation, where energy in the qubits dissipates as a result of interaction with the thermal environment, and dephasing, where the relative phase between quantum states starts randomising due to external EM fields, slow environmental changes or noise in the control systems. For these experiments, we do not consider errors associated with state preparation or measurement read-out. 

As stabiliser codes can only detect and correct combinations of $X$ and $Z$ errors, we only consider $X$, $Y$ and $Z$ errors for our noise models, in the form of probabilistic gate noise and depolarising noise. This means that our noise models are inherently unable to capture the full range of noise and errors that might arise in physical NISQ systems. However, since our aim is to evaluate the effectiveness of the [[4,2,2]] stabiliser code in improving training outcomes, we only need to simulate noise that the code is theoretically capable of detecting. 

We implement the gate noise model with single-qubit ``error" gates applied after each single-qubit gate, where there is a probability (or Pauli Error Rate), given by $p$ (with $0 < p < 1$), of either an $X$, $Y$ or $Z$ error occurring, and a $1-p$ chance of no error occurring. Additionally, we apply each single-qubit error gate after each 2-qubit gate, on the same qubits targeted by the 2-qubit gates. We also alter the Pauli Error Rate for 2-qubit gates so that it is double the error rate used for single-qubit gates, in order to better model the increased error rate for multi-qubit gates compared to single-qubit gates. 

Our environmental noise model is a highly simplified model that again consists of $X$,$Y$ and $Z$ errors. We inject Pauli errors into the system at regular intervals throughout the circuit, to each physical and ancilla qubit at the same time. Each injection has a Pauli Error Rate defined in the same way as for the gate noise model. Applying noise at regular intervals mimics the cumulative build-up of errors in quantum circuits that occur as a result of environmental noise, where the specific regularity of the noise injections reflects the typical relaxation time and dephasing time of the system. Although this model does not include amplitude damping noise, dephasing noise, or the entire span of complex errors that could arise from realistic noise models, it is able to capture a range of alterations that may occur to the qubits as a result of energy loss to the system and decoherence. The model is also compatible with our choice of QEC code. 

\section{Simulations}\label{sims}
We ran simulations of the logically-encoded circuit under the gate noise and environmental noise models in a classical high-performance computing environment, using Xanadu's Pennylane library~\cite{bergholm2022pennylaneautomaticdifferentiationhybrid} in \textsc{Python 3.12.3}. The application of each stabiliser and syndrome extraction adds one extra qubit to the system, hence the resource overhead for the simulations ranged from 12 qubits to 20 qubits depending on the number of syndrome extraction rounds performed. 

Since there are only four unique data points that can be used to train the VQC (namely, [0,0,0], [0,1,1], [1,0,1] and [1,1,0]), we duplicated the set to produce 40 training samples, and split it into 24 samples for training and 16 samples for testing. We used a batch size of 8 and ran the training for 100 iterations each, which was more than sufficient for convergence in a zero noise environment. Since the [[4,2,2]] stabiliser code is not capable of correcting errors, we discarded shots where at least one $X$ or $Z$ error was detected. Shots were rerun until no errors are detected. 

We apply the noise models to both ancilla and physical qubits in the system, but keep the syndrome extraction qubits noise-free. We chose a Pauli Error Rate ranging from $0.001$ to $0.01$ for both models, which is consistent with current NISQ device capabilities~\cite{Arute2019}. For the environmental noise model, we chose a Pauli error injection regularity of once every 4 gates, with the same Pauli Error Rates as used for the gate noise model. Since most of our gates are 2-qubit gates, with an estimated completion time of $\approx 10-200$ $ns$~\cite{Kubo_2023, Howard_2023, PhysRevApplied.22.024057}, and since our Pauli Error Rates produce one error every 100 to 1000 injections, equivalent to one error every 400 to 4000 gates, we estimate the time interval between errors to be $\approx 4 - 800$ $\mu s$. This is consistent with realistic relaxation and coherence times, which are roughly of order $10-1000$ $\mu s$ in superconducting qubits~\cite{PhysRevB.86.100506, PhysRevLett.130.267001}. 

Incoherent noise is normally modelled by mixed states and density matrices, as this formulation captures the true nature of how the quantum state evolves over time as a result of random noise. However, due to the large computational overhead of simulating the QML model training with density matrices (where size scales poorly with number of qubits), we modelled the system using statevectors instead. Although this model does not capture the decoherence of the original pure state, for our purposes it was able to correctly predict the probability vector output at the end of each circuit, while requiring much less computational time to simulate. 

\section{Results and Discussion}\label{Results}
We first present the impact of the noise models on training accuracy with and without error correction, focusing on the effectiveness of the [[4,2,2]] stabiliser code in detecting errors and protecting the training accuracies. We then show the impact of ancilla qubit noise on the fidelities of the physical qubit states and consequently, the training accuracies. We also reveal that ancilla qubit noise limits the effectiveness of the stabiliser code, and use our results to define a threshold for the maximum ancilla Pauli Error Rate and the minimum ancilla fidelity required for reliable error detection and best training accuracies. 

\subsection{Training with Noise without Error Detection}
\begin{figure*}
\centering
\includegraphics[trim={0cm 0cm 0cm 0.9cm}, clip=true, width=0.95\linewidth]{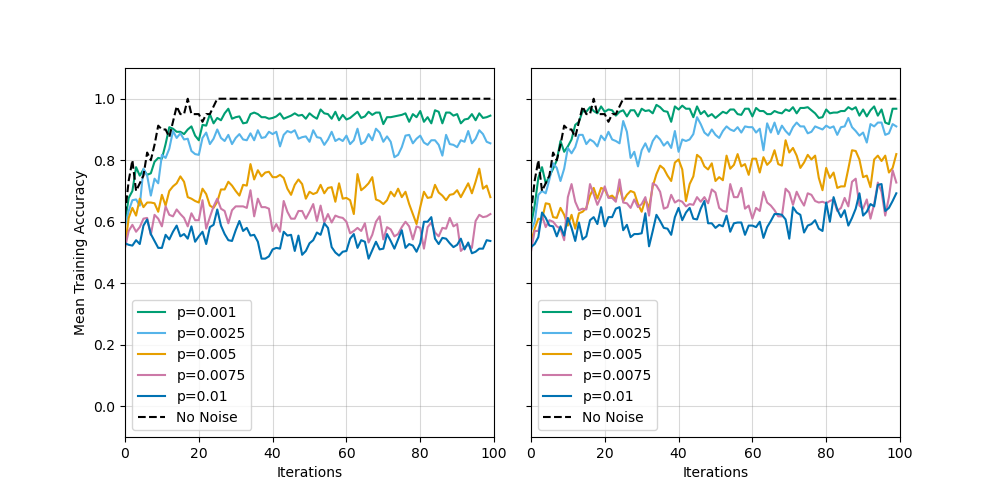}
\caption{\label{noerrordetection} Mean training accuracy of the logically-encoded VQC when training under different levels of gate (left) and environmental (right) noise, ranging from $p=0.001-0.01$. The black dashed line indicates the accuracy obtained from training without noise.}
\end{figure*}

In Figure~\ref{noerrordetection}, we show the evolution of the mean training accuracy achieved during training (where the mean was calculated from 10 simulations with different starting seeds), under both noise models, with varying noise levels expressed as Pauli Error Rates and without error detection. 

As we might expect, it is clear from Figure~\ref{noerrordetection} that the higher the noise level, the lower the final training accuracy is. For noise levels where $p\geq0.005$, the training curve does not exhibit a significant jump in accuracy within the first 30 iterations, as it does for the lower noise level simulations. Instead, they stay close to their initial training accuracy throughout training, indicating that the patterns in the data required for training are lost in the noise. When noise levels are at $p\leq0.0025$, learning appears hampered but not impossible. 

These results indicate that our simple VQC model is fairly susceptible to noise, which we suggest is due to its reliance on only two qubits to learn and make predictions. While larger and more complex QML algorithms generally have greater learning capacities and may be more robust to noise and errors, they also require more qubits and gates, which increases the potential for errors. Consequently, training and inference for both simple and complex QML algorithms on NISQ devices (where noise levels can exceed $p=0.005$) will most likely require a combination of error correction and mitigation. For some QML algorithms and applications, fully fault-tolerant quantum computation will still be necessary to achieve desirable levels of accuracy. 

\subsection{Training with Noise and Error Detection}
\begin{figure*}
\includegraphics[trim={0cm 0cm 0cm 0.9cm}, clip=true, width=0.95\linewidth]{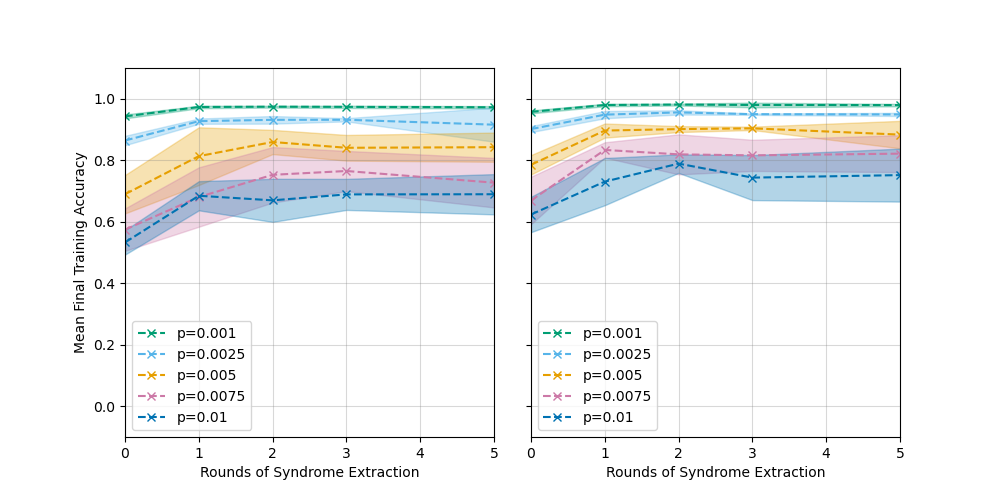}

\caption{\label{witherrordetection} Mean final training accuracy of the logically-encoded VQC under different levels of gate (left) and environmental (right) noise ranging from $p=0.001-0.01$, and with different degrees of error detection implemented from 0 to 5 rounds of syndrome extractions. The mean final training accuracy is calculated from the training accuracies attained by the VQC over the final 40 iterations of training, after convergence. The standard deviation of the mean final training accuracy from 10 training runs is represented by the shaded regions.}
\end{figure*}

In Figure~\ref{witherrordetection}, we display the effect of implementing the [[4,2,2]] stabiliser code with different numbers of syndrome extraction rounds on the final training accuracy achieved after model convergence. In this and subsequent figures, we are interested in the mean final training accuracy, which we calculated by taking the mean of the accuracies recorded over the last 40 iterations of training (as we can assume the training has stabilised by this point), and averaging this mean over 10 simulations of VQC training. We also report the first standard deviation associated with this mean. 

Under both gate noise and environmental noise models, we observe that for low noise levels (which we define as $p \leq 0.0025$), the training accuracy always improves with increasing number of syndrome extraction rounds and there is high consistency in the final training accuracy recorded across the 10 simulations. However, for higher levels of noise (specifically, $0.005 \leq p \leq 0.01$), the training accuracy is fairly inconsistent (greater spread in values), and occasionally worsens with more syndrome extractions. This runs counter to our expectation that more syndrome extractions should lead to the detection of more errors. Our results show that at higher noise levels, the effectiveness of syndrome extractions (and error detection) is limited. 

There is also evidence of limitations in the effectiveness of error detection at low noise levels. We find that even for $p\leq0.0025$, the increase in syndrome extractions produces no clear increase in final training accuracy beyond two extraction rounds, and does not reach $1.00$ even at low noise levels and with five rounds of syndrome extractions. 

These results suggest that there is noise in the logically-encoded circuit that syndrome measurements cannot detect. The only possible source of this noise is the ancilla qubits, which we do not apply any syndrome extractions to, but are entangled with the physical qubits via multiple CNOT gates. 

\begin{figure*}
\includegraphics[trim={0cm 0cm 0cm 0.8cm}, clip=true, width=\linewidth]{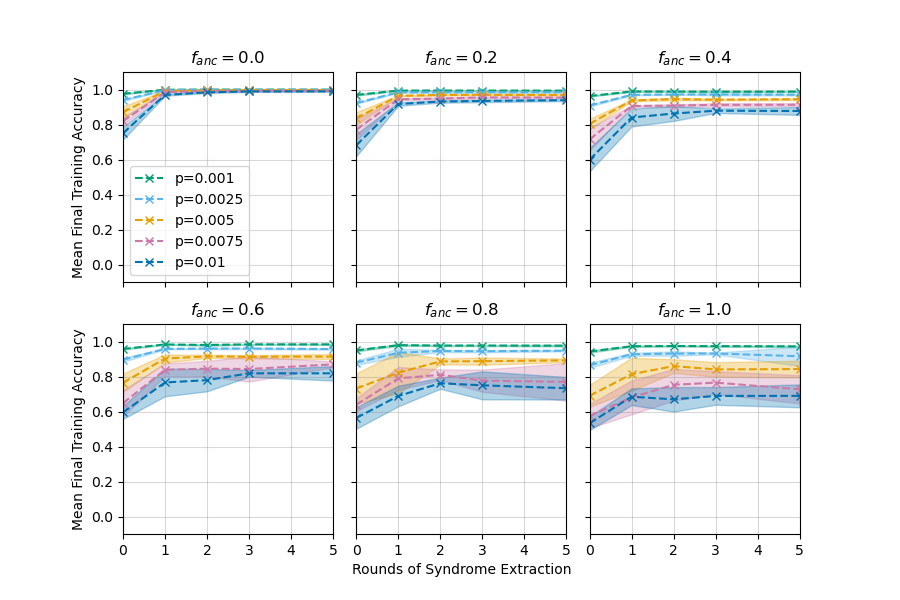}
\caption{\label{GateDP_Plot} Impact of ancilla qubit error rate on the mean final training accuracy, under the gate noise model with physical qubit error probability ranging from $p=0.001-0.01$. Each subplot displays the variation in mean final training accuracy with the number of rounds of syndrome extraction, for a specific combination of ancilla and physical qubit error probabilities. The ancilla error rates are expressed as a fraction of the physical error rates, denoted by the fraction $f_{anc}$. The standard deviation in the meaning final training accuracies, each calculated from 10 training runs, is shown by the shaded regions.}
\end{figure*}

\begin{figure*}
\includegraphics[trim={0cm 0cm 0cm 0.8cm}, clip=true, width=\linewidth]{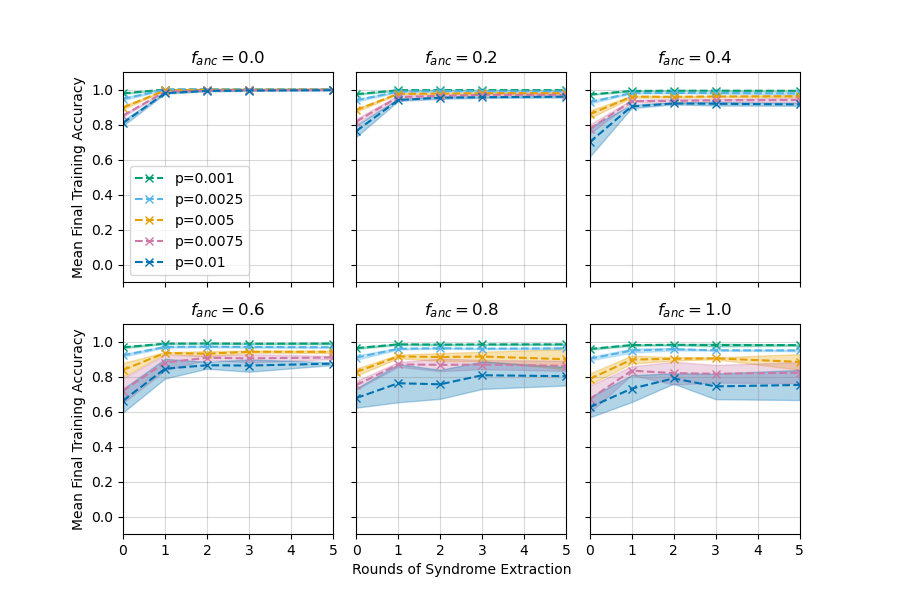}
\caption{\label{EN_plot} Impact of ancilla qubit error rate on the mean final training accuracy, under the environmental noise model with error probability ranging from $p=0.001-0.01$. Each subplot shows the variation in mean final training accuracy with the number of rounds of syndrome extraction, for a specific combination of ancilla and physical qubit error probabilities. The ancilla error rates are expressed as a fraction of the physical error rates, denoted by the fraction $f_{anc}$. The standard deviation in the meaning final training accuracies, each calculated from 10 training runs, is shown by the shaded regions.}
\end{figure*}

We ran the training with different levels of ancilla qubit noise, to determine the validity of our hypothesis that the ancilla qubit errors are responsible for the limited effectiveness of the [[4,2,2]] stabiliser code in detecting errors in the system. We show in Figures~\ref{GateDP_Plot} (gate noise model) and~\ref{EN_plot} (environmental noise model) the evolution in the mean final training accuracy as number of syndrome extractions increases, with different levels of ancilla qubit noise. The fraction $f_{anc}$ denotes the fraction of the physical qubit Pauli Error Rate that we apply to the ancilla qubits. For example, $f_{anc}=0.0$ means that there is no ancilla qubit noise, while $f_{anc}=0.5$ means that the ancilla Pauli Error Rate is half the physical Pauli Error Rate. 

It is clear from the plots that as the ancilla Pauli Error Rate increases, the syndrome extractions become less effective and less reliable, confirming our earlier hypothesis. There is a greater spread in the final training accuracy, as well as an increase in non-monotonicity (which we suggest is a consequence of the greater variance), when the ancilla error rate is higher. We note that lower noise levels produce better training outcomes than higher noise levels for $f_{anc} \neq 0.0$; specifically there is a far smaller chance of exhibiting non-monotonicity in the final training accuracy as the number of syndrome extractions increase, and a far smaller variance in the final training accuracy. Notably, when there is no noise on the ancilla qubits, the error detection works as expected and even at the highest noise levels, we see the training accuracy reach $1.0$ within five rounds of syndrome extractions. As long as there is any noise on the ancilla qubits, the error detection loses effectiveness, such that at higher ancilla noise levels (i.e., $p_{anc}\geq0.005$), the error detection becomes unreliable. 

Thus, a likely explanation for the high variability in final training accuracy at high ancilla qubit error rates is that the ancilla errors are not well-detected in this set-up, so they cannot be reliably eliminated from the training. We do not add syndrome extractions to the ancilla qubits, since it would require an additional encoding of the ancilla qubits, leading to more ancillary qubits that we cannot perform syndrome extractions on. Since the ancilla qubits are entangled with the physical qubits, their errors spread easily into the physical qubits through the CNOT gates (see Figure~\ref{logical_encoding}). While some of these errors will be detectable by syndrome measurements on the physical qubits, most of these errors will be non-Pauli. Hence, as the noise level increases, it becomes more difficult to protect the training through syndrome extractions, leading to the lower final accuracies and greater variance in its value. As we will demonstrate in Section~\ref{State_Fidelities}, high noise levels result in high spread in physical qubit state fidelities, which produces the higher variance and non-monotonicity in the final training accuracies. 

We again observe the tendency for the mean final training accuracies to plateau for $p_{anc} > 0$ instead of increase with more rounds of syndrome extraction. The plateauing effect is particularly apparent in the subplots of Figure~\ref{EN_plot} where ancilla qubits are subject to noise ($f_{anc} \neq 0$), but is most apparent where the ancilla Pauli Error Rate is less than approximately $0.003$. We also find that the higher the noise level, the lower the accuracy at which the plateau occurs. At higher levels of ancilla noise, the plateauing is hidden by the greater variance and non-monotonicity in the evolution of the final training accuracy. The plateau indicates that there is a limit for how many ancilla-caused errors can be detected and removed from the physical qubits by the stabiliser code for a given Pauli Error Rate, leaving only errors that syndrome measurements cannot detect. When that limit is reached, adding more syndrome extractions will not result in the detection of more errors, leading to the plateau. 

The plateauing is also present in the simulation results under gate noise (see Figure~\ref{GateDP_Plot}), though not as clearly visible because the threshold noise level before high variability takes over is lower than for the environmental noise simulations. The plateauing is most visible for $p_{anc} \approx 0.002 - 0.004$, where the ancilla noise is high enough to produce plateauing but low enough to not be masked by the high variability in final training accuracy. Interestingly, the plateauing starts at a higher number of syndrome extraction rounds in the simulations under the gate noise model than under the environmental noise model, suggesting that there are fewer detectable errors spreading to the physical qubits under the environmental noise model. Additionally, with the same ancilla and physical Pauli Error Rates, the gate noise model finishes at a lower final training accuracy than the environmental noise model, which reflects the difference in error levels between the two models. Given the greater frequency of errors produced and the higher error rate for 2-qubit gates under the gate model, it is unsurprising that the level of noise in the physical qubits, both detectable and undetectable, is greater in the gate noise model than in the environmental noise model. 

Our results for both noise models motivate the definition of a threshold Pauli Error Rate for ancilla qubits, such that when the error rate is larger than this threshold, error correction may not be effective - namely, the addition of more syndrome extractions may not improve training, there is high variability in final training accuracy, and the maximum mean final training accuracy is considerably lower than $1.0$ for the system. In determining the threshold Pauli Error Rate for our system, we excluded ancilla error rates that produce a plateau at a final training accuracy of less than $0.90$, even if high variability is not an issue. Taking into account all these requirements, we arrive at a threshold Pauli Error Rate of $p = 0.003$ for the gate noise model, and $p=0.004$ for the environmental depolarising noise model. For comparison, the current lowest single-qubit gate error rate exhibited by a NISQ device is $0.15\%$~\cite{Arute2019}, meaning that we may be able to run our simple VQC with the [[4,2,2]] code on the least noisy NISQ devices under special circumstances (for example, if the dominant noise is gate noise and environmental noise is very low). However, with the addition of real-world environmental noise to the gate noise, it is possible that the error threshold would be too low to run on currently available NISQ devices. 

There is a parallel between the threshold we have defined and the Threshold Theorem for quantum error correction, which asserts that there is a critical error rate below which sufficiently good quantum error correction codes can successfully correct errors. For error rates above this threshold, errors accumulate too quickly for effective error correction. However, despite the parallels, the Threshold Theorem does not explicitly cover the phenomenon of ancilla errors spreading to the physical qubits and reducing the effectiveness of the error correcting code. 

Though the threshold values we found are specific to our system and cannot be generalised to other systems, both the limit on the maximum training accuracy achievable and the existence of a threshold error rate for ancilla qubits should generalise to other combinations of QML algorithms and QEC codes. All QML algorithms contain rotation gates, and when implemented with QEC codes where ancilla qubits are needed for logically encoding rotation gates, resulting in the entanglement of ancilla and physical qubits, we can expect error propagation between the ancilla and physical qubit registers. If the QEC code cannot correct the full range of complex errors that may arise from such propagation, its effectiveness will be limited in noisy environments, leading to a maximum achievable training accuracy and a threshold error rate. Additionally, these results could potentially be relevant for implementing certain non-QML algorithms with QEC codes. This is because many QEC codes require ancilla qubits to logically encode rotation gates (and other non-Clifford gates), and since no known code can fully address the full spectrum of possible errors in practice, these implementations could also theoretically suffer from errors propagating from ancilla qubits to physical qubits. 

Our findings also highlight the need for additional considerations when applying techniques to achieve fault-tolerant implementations of QML algorithms. Fault-tolerant quantum machine learning may only be achievable with QEC if additional error mitigation techniques are used with it. For example, performing rotations with error mitigation instead of as logical operations (such as with zero-noise extrapolation~\cite{Pascuzzi_2022} or dynamical gate error correction~\cite{PhysRevA.80.032314}), and using flag qubits normally applied to syndrome qubits~\cite{PhysRevX.10.011022, PRXQuantum.1.010302} in the ancilla register to minimise errors before they propagate, may help minimise uncorrectable errors. 

\subsection{State Fidelities}
We next present our results showing the effect that noise and the implementation of error detection have on the state fidelities of physical and ancilla qubits, and use these results to further explain our training accuracy results above. All state fidelities ($F$) that we present in this subsection were calculated using the following definition: 
\begin{equation}
    F(\rho, \sigma) = |\langle \psi_{\rho} | \psi_{\sigma} \rangle|^2, 
\end{equation}
where $\rho$ and $\sigma$ represent the two pure quantum states we seek to compare (since we use pure states to represent the quantum states instead of mixed states). We use the error-free physical and ancilla states at the end of the logically-encoded circuit, just before measurement, as the ideal states to compare all other states to. 

We report only the absolute values of fidelities in this study. This is because we used probabilities to calculate expectation values during the training process, so any negative amplitudes just before measurement disappear and do not impact training. 

\subsubsection{State Fidelity Distributions}\label{State_Fidelities}
\begin{figure*}
\includegraphics[trim={0cm 0cm 0cm 1.0cm}, clip=true, width=0.9\linewidth]{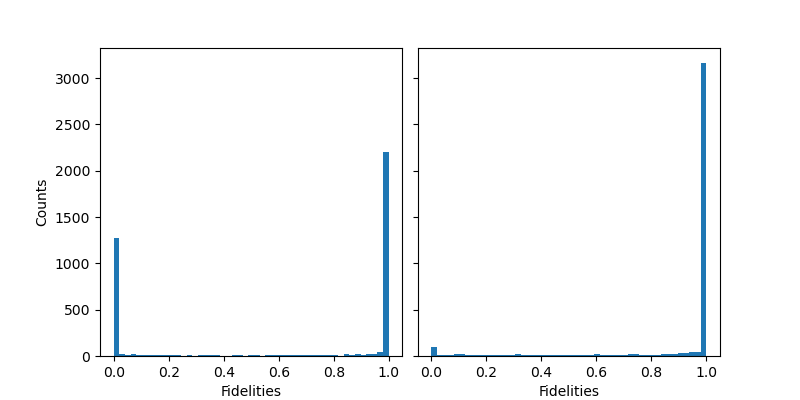}
\caption{\label{bimodal} The distribution of physical state fidelities displayed as histograms with bin size of 0.02, measured from 4000 training simulations, with a Pauli Error Rate of $p=0.01$ for physical and ancilla qubits. Left: Without error detection. Right: With error detection (three rounds of syndrome extractions).}
\end{figure*}

To determine state fidelity distributions under varying error rates and levels of error detection, we simulated the logically-encoded circuit with $4000$ shots, using each of the four possible inputs as initial states for $1000$ shots each. The analysis revealed roughly bimodal distributions, with peaks at 1.0 and 0.0 and a small proportion of intermediate fidelities\footnote{In the following discussions, we will occasionally use the Gaussian mean and standard deviation to characterise the fidelity distribution, as higher proportions of fidelities at or near 0 are sufficiently captured by the mean (through a reduction in the mean) and standard deviation (through an increase in the spread). It is this bimodality at the two extremes that produces the greater variability (and consequently, non-monotonicity) in the final training accuracies in Figures~\ref{GateDP_Plot} and~\ref{EN_plot} at the highest ancilla error rates, as higher error rates produce more (physical and ancilla) states with fidelities near 0. Example distributions for physical state fidelities with $p=0.01$ and $f_{anc}=1$ are illustrated in Figure~\ref{bimodal}.}. 

The distributions remain approximately bimodal for all error rates, with higher error rates producing more fidelities near 0 (and between 0 and 1), and fewer fidelities close to 1. As shown in Figure~\ref{bimodal}, applying error detection improves the proportion of fidelities near 1 by removing the states with errors from the distribution. The near-bimodality of the distributions suggests that errors in physical or ancilla qubits will often yield states orthogonal to the correct state, which we suggest is due to the nature of the errors consisting of only bitflips and phaseflips. Bitflips in particular are far more likely to create orthogonal states from flipping at least one of the qubits in the system. For example, if the physical qubit state was originally forming the logical $\ket{00}_{L}$ state, the only non-orthogonal state that can result from bitflips is the original state itself. Additionally, rotation gates will preserve the orthogonality of the errored state relative to the correct state, as rotations are unitary operations. Phase flips will have a smaller and less predictable impact on state fidelities, but in theory could create both orthogonal and non-orthogonal states, adding to the number of states with fidelities near 0 and introducing states with fidelities between 0 and 1. Both bitflips and phase flips occurring on ancilla qubits will spread to the physical qubits without necessarily preserving orthogonality, leading to physical qubit states with intermediate fidelities (and vice versa). 

\subsubsection{Pauli Error Rates and State Fidelities}
Table~\ref{fidelities_trainingaccuracies} presents the mean fidelities (and standard deviations) for ancilla and physical qubits under differing Pauli Error Rates and without error correction, calculated from the previously mentioned $4000$-shot simulations. Since the fidelity distributions are non-Gaussian, we report both the mean (and standard deviation\footnote{The standard deviations imply fidelity ranges that exceed the maximum value of 1, which is due to the non-Gaussianity of the fidelity distributions.}) and the fraction of fidelities below $0.02$ or above $0.98$. Additionally, we provide the corresponding mean final training accuracies (with standard deviations) calculated from 10 independent training runs, each initiated with a unique random seed. 

\begin{table*}
\centering
\begin{tabular}{c|c|c|c|c|c|c|c|c|c|c|c|c}
    \hline
     Noise & $p_{phys}$ & $f_{anc}$ & $p_{anc}$ & $F_{anc}$ & $F_{anc}$ & $F_{anc}<0.02$ & $F_{anc}>0.98$ & $F_{phys}$ & $F_{phys}$ & $F_{phys}<0.02$ & $F_{phys}>0.98$ & Training\\
     Model &  & &  & Mean & Std & fraction & fraction & Mean & Std & fraction & fraction & Accuracy\\
     \hline
     \hline
     Gate & 0.001 & 0.0 & 0.0 & $0.99$ & $0.07$ & 0.00 & 0.99 & $0.96$ & $0.18$ & 0.03 & 0.96 & $0.975 \pm 0.004$ \\
     Noise & & 0.2 & 0.0002 & $0.98$ & $0.12$ & 0.01 & 0.97 & $0.96$ & $0.19$ & 0.03 & 0.95 & $0.969 \pm 0.004$\\
     & & 0.5 & 0.0005 & $0.97$ & $0.17$& 0.02 & 0.96 & $0.95$ & $0.19$ & 0.03 & 0.94 & $0.960 \pm 0.006$\\
     & & 0.8 & 0.0008 & $0.95$ & $0.20$ & 0.03 & 0.93 & $0.94$ & $0.21$ & 0.04 & 0.93 & $0.949 \pm 0.006$\\
     & & 1.0 & 0.001 & $0.93$ & $0.25$ & 0.05 & 0.91 & $0.94$ & $0.22$ & 0.04 & 0.92 & $0.943 \pm 0.006$\\
     \cline{2-13}
     & 0.005 & 0.0 & 0.0 & $0.97$ & $0.13$ & 0.00 & 0.94 & $0.83$ & $0.37$ & 0.15 & 0.81 & $0.87 \pm 0.03$\\
     & & 0.2 & 0.001 & $0.92$ & $0.25$ & 0.05 & 0.87 & $0.82$ & $0.37$ & 0.15 & 0.79 & $0.83 \pm 0.04$\\
     & & 0.5 & 0.0025 & $0.83$ & $0.35$ & 0.12 & 0.78 & $0.79$ & $0.39$ & 0.17 & 0.74 & $0.80 \pm 0.02$\\
     & & 0.8 & 0.004 & $0.75$ & $0.41$ & 0.18 & 0.69 & $0.75$ & $0.41$ & 0.18 & 0.69 & $0.73 \pm 0.09$\\
     & & 1.0 & 0.005 & $0.71$ & $0.43$ & 0.22 & 0.64 & $0.75$ & $0.41$ & 0.18 & 0.67& $0.69 \pm 0.06$\\
     \cline{2-13}
     & 0.010 & 0.0 & 0.0 & $0.95$ & $0.18$ & 0.01 & 0.90 & $0.71$ & $0.44$ & 0.25 & 0.68 & $0.75 \pm 0.04$\\
     & & 0.1 & 0.001 & $0.89$ & $0.28$ & 0.25 & 0.65 & $0.70$ & $0.44$ & 0.25 & 0.66 & $0.70 \pm 0.06$\\
     & & 0.4 & 0.004 & $0.74$ & $0.41$ & 0.19 & 0.66 & $0.64$ & $0.45$ & 0.28 & 0.57 & $0.60 \pm 0.07$\\
     & & 0.5 & 0.005 & $0.70$ & $0.43$ & 0.22 & 0.62 & $0.63$ & $0.46$ & 0.29 & 0.55 & $0.58 \pm 0.05$\\
     & & 1.0 & 0.01 & $0.52$ & $0.47$ & 0.38 & 0.43 & $0.57$ & $0.46$ & 0.31 & 0.46 &$0.53 \pm0.04$\\
     \hline
     \hline
     Environ. & 0.001 & 0.0 & 0.0 & $0.99$ & $0.05$ & 0.00 & 0.99 & $0.97$ & $0.17$ & 0.03 & 0.97 & $0.978 \pm 0.004$\\
     Noise & & 0.2 & 0.0002 & $0.98$ & $0.11$ & 0.01 & 0.98 & $0.96$ & $0.17$ & 0.03 & 0.96 & $0.973 \pm 0.004$\\
      & & 0.5 & 0.0005 & $0.98$ & $0.15$ & 0.02 & 0.97 & $0.96$ & $0.19$ & 0.03 & 0.95 & $0.969\pm 0.003$\\
     & & 0.8 & 0.0008 & $0.96$ & $0.19$ & 0.03 & 0.95 & $0.95$& $0.19$ & 0.03 & 0.94 & $0.962 \pm 0.005$\\
     & & 1.0 & 0.001 & $0.95$ & $0.20$ & 0.04 & 0.94 & $0.95$ & $0.19$ & 0.03 & 0.94 & $0.957 \pm 0.005$\\
     \cline{2-13}
     & 0.005 & 0.0 & 0.0 & $0.97$ & $0.13$ & 0.01 & 0.95 & $0.85$ & $0.35$ & 0.14 & 0.84 & $0.90 \pm 0.01$\\
     & & 0.2 & 0.001 & $0.93$ & $0.23$ & 0.04 & 0.90 & $0.84$ & $0.35$ & 0.14 & 0.82 & $0.88 \pm 0.01$\\
     & & 0.5 & 0.0025 & $0.89$ & $0.30$ & 0.08 & 0.85 & $0.83$ & $0.36$ & 0.14 & 0.79& $0.82 \pm 0.07$\\
     & & 0.8 & 0.004 & $0.82$ & $0.37$ & 0.14 & 0.77 & $0.82$ & $0.36$ & 0.14 & 0.76 & $0.83 \pm 0.02$\\
     & & 1.0 & 0.005 & $0.78$ & $0.40$ & 0.18 & 0.74 & $0.80$ & $0.37$ & 0.14 & 0.74 & $0.79 \pm 0.03$\\
     \cline{2-13}
     & 0.010 & 0.0 & 0.0 & $0.95$ & $0.17$ & 0.01 & 0.91 & $0.72$ & $0.44$ & 0.25 & 0.70 & $0.81 \pm 0.02$\\
     & & 0.1 & 0.001 & $0.91$ & $0.26$ & 0.04 & 0.86 & $0.71$ & $0.44$ & 0.25 & 0.68 & $0.79 \pm 0.02$\\
     & & 0.4 & 0.004 & $0.79$ & $0.38$ & 0.16 & 0.73 & $0.70$ & $0.44$ & 0.24 & 0.64 & $0.70 \pm 0.08$\\
     & & 0.5 & 0.005 & $0.77$ & $0.40$ & 0.18 & 0.71 & $0.69$ & $0.44$ & 0.25 & 0.62 &$0.66 \pm 0.09$\\
     & & 1.0 & 0.01 & $0.60$ & $0.47$ & 0.33 & 0.54 & $0.65$ & $0.45$ & 0.27 & 0.55 & $0.62 \pm 0.06$\\
     \hline
     \hline
\end{tabular}
\caption{Ancilla qubit fidelities and physical qubit fidelities under selected ancilla and physical qubit Pauli Error Rates, and without error detection. Fidelities are reported as means along with standard deviations, and in terms of the fraction of fidelities under $0.02$ and over $0.98$. Training accuracies are reported as means with error bounds given by standard deviations.}
\label{fidelities_trainingaccuracies}
\end{table*}

Our results reveal several noteworthy trends and observations. Firstly, ancilla and physical qubit fidelities are primarily affected by their own Pauli Error Rates, but are slightly impacted by the error rate of the other qubit register. For example, when the physical error rate is non-zero and the ancilla error rate is zero, the ancilla fidelities still fall below $1.0$, as physical qubit errors can propagate to ancilla qubits through the CNOT gates. We observe this also in the gate noise model, when the physical and ancilla error rate is $0.001$, the proportion of fidelities where $F_{anc} > 0.98$ is $0.91$ (with $\bar{F}_{anc} = 0.93$), but when the physical error rate is increased to $0.010$ while keeping ancilla error rate the same, the proportion of fidelities with $F_{anc} > 0.98$ is reduced to $0.65$ (with $\bar{F}_{anc} = 0.89$). Increasing the ancilla error fraction also reduces the physical state fidelities, despite no change in physical error rates. 

Secondly, the mean final training accuracy is not especially robust to reductions in the physical state fidelities, which matches with our earlier observation that the training accuracies are impacted by even low levels of noise. To achieve an accuracy above 0.95, the mean physical state fidelity needs to be at or higher than roughly 0.95\footnote{The fidelity requirements for reaching high accuracies are dependent on the model and application; some implementations may not require such high fidelities.}, corresponding to approximately $93-94\%$ of states with $F > 0.98$ and $3-4\%$ with $F < 0.02$. Apart from reducing the mean training accuracy, errors will also increase its standard deviation by increasing the number of very low fidelity states that occur during training. The training accuracies follow a much more Gaussian distribution than the fidelities, as many states are used to train the model, which tends to average out the impact of errors on training. 

We expect to find that under both noise models, the training accuracy should follow the physical state fidelities at the end of the circuit very closely, as it is the physical state that is fed to the algorithm for optimisation. However, our results slightly deviate from this expectation. For example, in the environmental noise model, for $p_{phys} = 0.005$, we find that when $p_{anc} = 0.0025$, the physical state fidelity is $0.89$ and the mean final training accuracy is $0.82 \pm 0.07$, whereas when $p_{anc}=0.004$, the physical state fidelity is $0.82$ and the mean final training accuracy is $0.83 \pm 0.02$. The similarity in final training accuracy despite clear difference in physical state fidelity is most likely from using only 10 samples to determine mean training accuracy and standard deviation. 

Finally, we note that for both noise models, ancilla fidelities are lower than physical fidelities at the same error rates. This is because there are more gate operations applied to the ancilla register than physical register, and thus a higher chance of both gate errors and a greater level of simulated environmental noise in the ancilla register. The ancilla register thus accumulates more errors than the physical register under both noise models. 

\subsubsection{Error Detection and State Fidelities}
\begin{figure*}
\includegraphics[trim={1.3cm 0.4cm 1.6cm 1.2cm}, clip=true, width=0.9\linewidth]{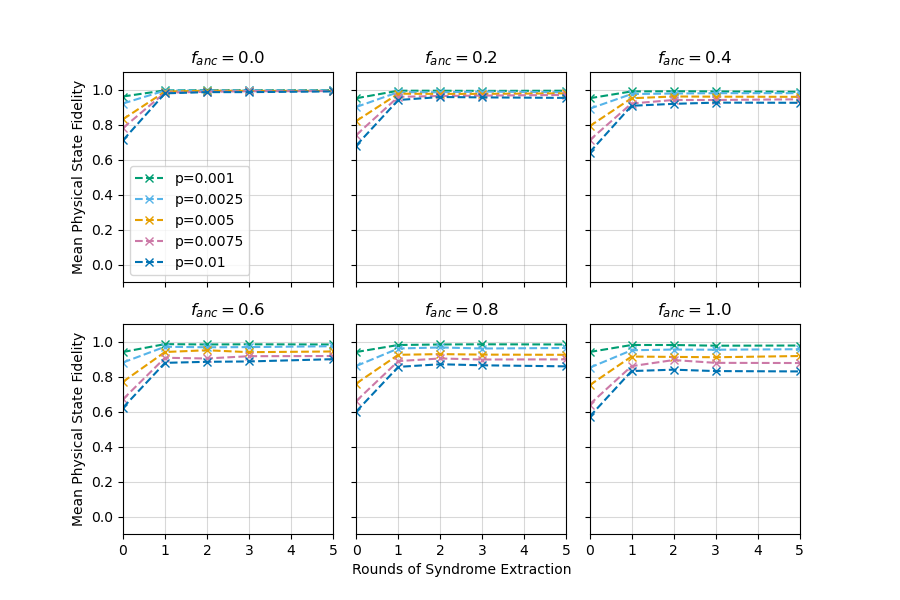}
\caption{\label{phys_fids_with_correction} Mean physical state fidelity as a function of number of syndrome extraction rounds under the gate noise model, for different rates of ancilla error. The means are calculated from 10 training runs per combination of ancilla and physical error rate, where the physical error rate ranges from $p=0.001-0.01$, and the ancilla error rate is expressed as a fraction of the physical error rates and denoted by $f_{anc}$. The standard deviations in fidelity are not displayed to ensure visibility.}
\end{figure*}

\begin{figure*}
\includegraphics[trim={1.3cm 0.4cm 1.6cm 1.2cm}, clip=true, width=0.9\linewidth]{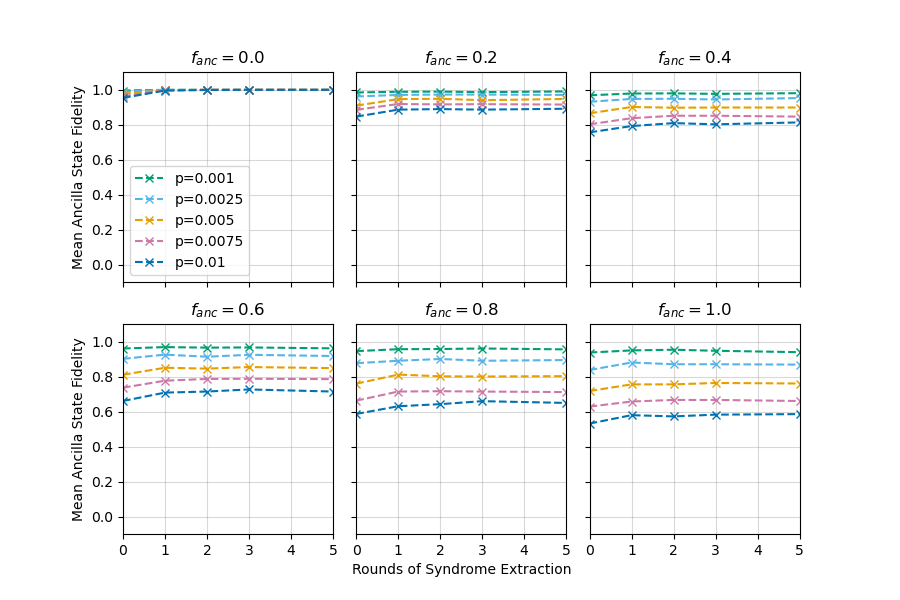}
\caption{\label{anc_fids_with_correction} Mean ancilla state fidelity as a function of number of syndrome extraction rounds under the gate noise model, for different rates of ancilla error. The means are calculated from 10 training runs per combination of ancilla and physical error rate, where the physical error rate ranges from $p=0.001-0.01$, and the ancilla error rate is expressed as a fraction of the physical error rates and denoted by $f_{anc}$. The standard deviations in fidelity are not displayed to ensure visibility.}
\end{figure*}

We show the impact of error detection on the mean physical and ancilla fidelities in Figures~\ref{phys_fids_with_correction} and~\ref{anc_fids_with_correction}, respectively, under the gate noise model and with varying Pauli Error Rates on both registers. While there is slight improvement in ancilla fidelity after one round of syndrome extractions, likely from removing the errors that would have otherwise propagated from the physical to ancilla register, the application of more syndrome extractions to the physical qubits has minimal effect on ancilla fidelities. Additionally, we notice that the physical fidelities also plateau after approximately one set of syndrome extractions, and lower mean ancilla fidelities are associated with lower mean physical fidelities, at their respective plateaus. These observations support our earlier assertion that ancilla errors impact the physical qubits in ways that cannot be detected by the [[4,2,2]] stabiliser code. More fundamentally, the Pauli errors that originate in the ancilla register and are propagated to the physical register through entangling gates may not remain Pauli errors when they reach the physical qubits. It is these non-Pauli errors that the stabiliser code is unable to detect effectively, resulting in a maximum mean physical qubit fidelity that can be reached even with error detection, and consequently a maximum mean training accuracy. We also observe very similar trends under the environmental noise model. 

We identified the ancilla fidelities corresponding to the threshold Pauli Error Rates we established earlier: $p_{anc} = 0.003$ and $p_{anc} = 0.004$. Since ancilla fidelities tend to stabilise and plateau after one round of syndrome extractions, we determined the threshold ancilla fidelities by averaging the fidelity measurements taken at 1, 2, 3, and 5 rounds of syndrome extractions, which should all be close to the true ancilla fidelity at the threshold error rate. Table \ref{fidelities_at_threshold} reports the mean ancilla and physical fidelities we measured in this way from the 4000-shot simulations, at ancilla error rates slightly above, equal to, and slightly below the thresholds. We included some ancilla error rates more than once, with different physical Pauli Error Rates. We generally find that the ancilla fidelities remain roughly the same despite differences in the physical Pauli Error Rates. This is because after one round of syndrome extractions is applied, the vast bulk of the impact from the errors in the physical register is removed, leaving non-Pauli errors that cannot be removed. We conclude from our results that the thresholds for mean ancilla fidelities that correspond to the Pauli Error Rate thresholds are $0.85$ and $0.83$, for the gate noise model and the environmental noise model, respectively. The corresponding proportion of states with fidelities near 0 and 1 at the thresholds are $12\%$ and $82\%$ for the gate noise model, and $14\%$ and $82\%$ for the environmental noise model. 

\begin{table*}
\centering
\begin{tabular}{c|c|c|c|c|c|c|c|c|c}
    \hline
     Noise & $p_{phys}$ & $f_{anc}$ & $p_{anc}$ & Mean & $F_{anc}<0.02$ & $F_{anc}>0.98$ & Mean & $F_{phys}<0.02$ & $F_{phys}>0.98$ \\
     Model &  &  &  & $F_{anc}$ & fraction & fraction & $F_{phys}$& fraction & fraction \\
     \hline
     \hline
     Gate Noise & 0.0100 & 0.2 & 0.002 & 0.89 & 0.09 & 0.87 & 0.95 & 0.01 & 0.92 \\
     & 0.0025 & 1.0 & 0.0025 & 0.87 & 0.10 & 0.85 & 0.96 & 0.01 & 0.92 \\
     & 0.0050 & 0.6 & \textbf{0.003} & \textbf{0.85} & 0.12 & 0.82 & 0.96 & 0.02 & 0.85\\
     & 0.0075 & 0.4 & \textbf{0.003} & \textbf{0.85} & 0.12 & 0.82 & 0.94 & 0.01 & 0.90 \\
     & 0.0100 & 0.4 & 0.004 & 0.81 & 0.15 & 0.78 & 0.92 & 0.02 & 0.86 \\
     & 0.0075 & 0.6 & 0.0045 & 0.79 & 0.17 & 0.75 & 0.92 & 0.02 & 0.85 \\
     \cline{2-10}
     \hline
     \hline
     Environ. & 0.005 & 0.6 & 0.003 & 0.87 & 0.11 & 0.85 & 0.96 & 0.01 & 0.93\\
     Noise & 0.0075 & 0.4 & 0.003 & 0.88 & 0.10 & 0.86 & 0.96 & 0.01 & 0.92\\
      & 0.010 & 0.4 & \textbf{0.004} & \textbf{0.83} & 0.14 & 0.81 & 0.94 & 0.01 & 0.89\\
      & 0.005 & 0.8 & \textbf{0.004} & \textbf{0.83} & 0.14 & 0.82 & 0.95 & 0.01 & 0.90\\
      & 0.0075 & 0.6 & 0.0045 & 0.81 & 0.16 & 0.79 & 0.94 & 0.01 & 0.89\\
     & 0.005 & 1.0 & 0.005 & 0.80 & 0.17 & 0.78 & 0.94 & 0.01 & 0.88\\
     \cline{2-10}
     \hline
     \hline
\end{tabular}
\caption{Ancilla qubit fidelities ($F_{anc}$) and physical qubit fidelities ($F_{phys}$) for selected ancilla ($p_{anc}$) and physical ($p_{phys}$) qubit Pauli Error Rates, with threshold ancilla fidelities highlighted in bold text. Fidelities are reported in terms of the mean and proportions with $F < 0.02$ and $F > 0.98$.}
\label{fidelities_at_threshold}
\end{table*}

It is important to highlight that these threshold fidelities and Pauli Error Rates are dependent on the models we have used in this study. For different noise models and a different VQC, we cannot guarantee that the threshold fidelities and error rates will remain in the same ballpark. However, our insights about how errors interact and propagate between ancilla and physical registers, their limiting effect on error correction schemes and the existence of a threshold error rate for ancilla qubits (and associated ancilla fidelity) are highly relevant to the general implementation of VQCs with error correcting codes on both NISQ devices and fault-tolerant devices. Any QEC implementation that relies on ancilla qubits for encoding and is limited in the type of error it can detect and correct, will have a limiting error rate for which error correction cannot effectively protect the training and prediction processes. This limit is in addition to the limit accounted for by the Threshold Theorem. 

\section{Conclusions}\label{conclusion}
Through classical simulations of a 2-qubit VQC implemented with the [[4,2,2]] code, we have demonstrated that applying QEC can improve QML training accuracies in noisy environments, proving that even in a non-fault tolerant setting, QEC is useful for practical QML implementations. However, the effectiveness of error correction is limited by the error rate of ancilla qubits. We can define a threshold ancilla error rate such that the QEC code can reliably guarantee a reasonable final training accuracy if the ancilla error rate is below the threshold, and such that above the threshold, training accuracies may be poor and quite variable. 

Under our gate noise and environmental noise models, respectively, we determined this threshold to be $p_{anc}=0.003$ and $p_{anc}=0.004$, for a desired minimum training accuracy of $0.90$. The ancilla fidelities corresponding to these error rates are $0.85$ and $0.83$, respectively. The threshold error rate for the gate noise model compares favourably to the lowest single-qubit gate error rates exhibited by a NISQ device of $0.15\%$. Under a more complex noise model with both gate noise and environmental noise, we would likely see a lower threshold required for the stabiliser code to detect errors effectively, as a greater number of errors would propagate from ancilla to physical registers under such a noise model. Moreover, both models are simplified models of realistic gate noise and environmental noise. Including other forms of environmental noise-induced errors such as amplitude damping, or non-Pauli errors from rotation gates, would likely lead to lower threshold error rates, potentially lower than the lowest error rates exhibited by current NISQ systems. 

Our proposed explanation for the observed threshold is the propagation of Pauli errors from ancilla qubits to physical qubits through the combination of CNOT gates and rotation gates, which transform the Pauli errors into non-Pauli errors that are undetectable by the stabiliser code. Although our results come from a single system consisting of a 2-qubit VQC and the [[4,2,2]] error-detecting code, the physical phenomena that give rise to these results are clearly generalisable to other combinations of QML and QEC algorithms. We conclude from our results that any QML algorithm and QEC code implementation where rotation gates need ancilla qubits to logically encode non-transversally, will allow errors to propagate between ancilla and physical qubits, leading to the formation of exotic errors in the physical qubits. If the QEC code employed is not able to detect or correct for these types of errors, the effectiveness of the QEC code is significantly hampered at high error rates and training accuracies achievable by the QML system will also be limited. The specific limit on achievable accuracy depends on the ancilla error rate and the capabilities of the QEC code. 

These findings indicate that practical implementation of QML algorithms on NISQ systems requires consideration of both the logical encoding associated with the QEC code and the code's capacity to detect a wide range of error types, in addition to error mitigation approaches to be used in conjunction with the QEC code. Given the limitations of a purely QEC-based approach, it is worth exploring alternative methods to use in addition to QEC codes. For example, flag qubits may be employed to address errors on the ancilla qubits before they have a chance to propagate to the physical register. If errors propagating between ancilla and physical qubits cannot be fully corrected or detected, we cannot perform QML algorithms with logical error rates suppressed to arbitrarily low levels even with fully fault-tolerant systems. 

\begin{acknowledgments}
The authors acknowledge the support of the CSIRO Impossible Without You Program. This research/project relied on the use of resources and services from the CSIRO’s High Performance Computing cluster (Petrichor). 
\end{acknowledgments}

\bibliography{EAdermann_VQC_QEC}

\end{document}